\documentclass[preprint]{elsarticle}
\usepackage{graphicx}
\usepackage{txfonts}
\usepackage{color}
\usepackage{natbib}
\usepackage{tabularx}
\usepackage{lineno}
\usepackage{multirow}
\usepackage{multicol}
\usepackage{slashbox}
\bibpunct{(}{)}{;}{a}{}{,} 

\begin{document}

\begin{frontmatter}

\title{Effect of O$_3$ on the atmospheric temperature structure of early Mars}

\author[lab,cnrs]{P. von Paris\corref{cor}}
\author[lab,cnrs]{F. Selsis}
\author[dlr]{M. Godolt}
\author[dlr]{J.L. Grenfell}
\author[dlr,zaa]{H. Rauer}
\author[dlr]{B. Stracke}

\cortext[cor]{Corresponding author: email vonparis@obs.u-bordeaux1.fr, tel. +33 (0)557 77 6131}

\address[lab]{Univ. Bordeaux, LAB, UMR 5804, F-33270, Floirac, France}
\address[cnrs]{CNRS, LAB, UMR 5804, F-33270, Floirac, France }
\address[dlr]{Institut f\"{u}r Planetenforschung, Deutsches Zentrum
f\"{u}r Luft- und Raumfahrt (DLR), Rutherfordstr. 2, 12489 Berlin, Germany}
\address[zaa]{Zentrum f\"{u}r Astronomie und Astrophysik (ZAA), Technische
Universit\"{a}t Berlin, Hardenbergstr. 36, 10623 Berlin, Germany}

\begin{abstract}

Ozone is an important radiative trace gas in the Earth's atmosphere and has also been detected on Venus and Mars. The presence of ozone can significantly influence the thermal structure of an atmosphere due to absorption of stellar UV radiation, and by this e.g. cloud formation. Photochemical studies suggest that ozone can form in carbon dioxide-rich atmospheres. Therefore, we investigate the effect of ozone on the temperature structure of simulated early Martian atmospheres.

With a 1D radiative-convective model, we calculate temperature-pressure profiles for a 1\,bar carbon dioxide atmosphere containing various amounts of ozone. These ozone profiles are fixed, parameterized profiles. We vary the location of the ozone layer maximum and the concentration at this maximum. The maximum is placed at different pressure levels in the upper and middle atmosphere (1-10\,mbar). 

Results suggest that the impact of ozone on surface temperatures is relatively small. However, the planetary albedo significantly decreases at large ozone concentrations. Throughout the middle and upper atmospheres, temperatures increase upon introducing ozone due to strong UV absorption. This heating of the middle atmosphere strongly reduces the zone of carbon dioxide condensation, hence the potential formation of carbon dioxide clouds. For high ozone concentrations, the formation of carbon dioxide clouds is inhibited in the entire atmosphere.  In addition, due to the heating of the middle atmosphere, the cold trap is located at increasingly higher pressures when increasing ozone. This leads to wetter stratospheres hence might increase water loss rates on early Mars. However, increased stratospheric H$_2$O would lead to more HO$_x$, which could efficiently destroy ozone by catalytic cycles, essentially self-limiting the increase of ozone. This result emphasizes the need for consistent climate-chemistry calculations to assess the feedback between temperature structure, water content and ozone chemistry. Furthermore, convection is inhibited at high ozone amounts, leading to a stably stratified atmosphere. 

\end{abstract}

\end{frontmatter}

\section{Introduction}

The climate and, consequently, the habitability of early Mars are a long-standing question that has not yet been answered conclusively. From an atmospheric modeling point of view, short episodes of a warm, wet climate, followed by long periods of arid, cold climates, seem to be favored for the Noachian period about 3.8 billion years ago (e.g., \citealp{segura2008}, \citealp{wordsworth2013}, \citealp{halevy2014}).

An important input for atmospheric modeling studies is the atmospheric composition, together with the surface pressure. Most previous studies (e.g., \citealp{kasting1991}, \citealp{wordsworth2013}) assumed mixed carbon dioxide (CO$_2$)-water vapor (H$_2$O) atmospheres. In these studies, the partial pressure of carbon dioxide was of the order of bars, consistent with outgassing model studies (e.g., \citealp{phillips2001}, \citealp{grott2011}) and in-situ analyses (e.g., \citealp{manga2012}, \citealp{kite2014}). In 1D models, the atmospheric H$_2$O content is usually controlled by a fixed relative humidity to simulate a hydrological cycle. 

However, other atmospheric (trace) gases such as molecular nitrogen (N$_2$), molecular hydrogen (H$_2$), methane (CH$_4$) or sulphur dioxide (SO$_2$) might have been present in the early Mars atmosphere. The initial N$_2$ inventory is thought to be relatively large (e.g., \citealp{mckay1989mars}) and can provide modest surface warming \citep{vparis2013marsn2}. Methane has also been suggested to warm the atmosphere (e.g., \citealp{postawko1986}, \citealp{ramirez2014}), but the effect on surface temperature is small. \citet{ramirez2014} find that H$_2$-induced warming could raise surface temperatures above freezing, assuming H$_2$ was a major atmospheric constituent (around 20\,\% volume mixing ratio). The effect of SO$_2$ has been investigated by a number of studies (e.g., \citealp{postawko1986}, \citealp{yung1997}, \citealp{halevy2007}, \citealp{johnson2008}, \citealp{tian2010}, \citealp{mischna2013}), but results suggest that it most likely does not contribute strongly to warming the surface.

One proposed solution to warming early Mars has been the formation of carbon dioxide clouds (e.g., \citealp{pierre1998}, \citealp{forget1997}). Early 1D modeling studies suggested that the needed warming however strongly depends on the assumed cloud cover, which would have to be nearly 100\% (e.g., \citealp{mischna2000}). Following time-dependent 1D modeling studies by \citet{Cola2003} found this to be unrealistic, a conclusion also supported by recent 3D studies (e.g., \citealp{forget2013}, \citealp{wordsworth2013}). In addition, \citet{kitzmann2013} suggested that previous radiative transfer algorithms probably strongly overestimated the warming associated with carbon dioxide clouds.

The formation of carbon dioxide clouds depends on the temperature profile in the middle atmosphere. As shown by, e.g., \citet{yung1997} or \citet{ramirez2014}, UV absorption by SO$_2$ or near-IR absorption by CH$_4$ can effectively inhibit carbon dioxide condensation, or at least reduce the cloud formation region.

So far, ozone (O$_3$) has not been considered in the context of early Mars. Like SO$_2$, ozone has strong UV and visible absorption bands. On Earth, these are responsible for the pronounced stratospheric temperature maximum. Ozone has been detected in the CO$_2$-rich atmospheres of both Venus (e.g., \citealp{montmessin2011}) and Mars (e.g., \citealp{lebonnois2006}). Furthermore, numerous photochemical modeling studies suggests that ozone can be formed from abiotically produced oxygen (e.g., \citealp{selsis2002}, \citealp{segura2007}, \citealp{domagal2010,domagal2014}). In addition, \citet{wordsworth2014} propose a (non-chemical) abiotic oxygen formation process based on hydrogen escape and water cold-trapping. Therefore, it is reasonable to assume that on early Mars, ozone could have been also, to a certain extent, present in the atmosphere. Hence, in this study, we explore the effect of ozone on the temperature structure of early Mars 

The paper is structured as follows: Section \ref{model} presents the atmospheric model and the simulations performed. Results are shown and discussed in Sect. \ref{results} and conclusions in Sect. \ref{summary}.

\section{Computational details}

\label{model}

\subsection{Model description}

We use a 1D, steady-state, cloud-free radiative-convective atmosphere model to calculate globally, diurnally averaged temperature and H$_2$O profiles. The model is originally based on \citet{kasting1984water} and \citet{kasting1984}. Further code developments are described in, e.g., \citet{kasting1988}, \citet{mischna2000}, \citet{vparis2008} or \citet{vparis2010gliese}. 

The model atmospheres are divided into 52 levels. Temperature profiles in the upper atmosphere are obtained by solving the radiative transfer equation (38 bands for incoming stellar radiation, 25 bands for planetary and atmospheric thermal radiation). For incoming stellar radiation (0.2-4.5\,$\mu$m), Rayleigh scattering (by N$_2$, H$_2$O, CO$_2$ and CH$_4$) and molecular absorption (by H$_2$O, CO$_2$, CH$_4$ and O$_3$) contribute to the opacity. To allow for scattering, the angular integration of the radiative transfer equation is performed using a 2-stream code \citep{Toon1989}. For thermal radiation (1-500\,$\mu$m), molecular absorption (by H$_2$O, CO$_2$, CH$_4$ and O$_3$, see below, Sect. \ref{improv}) and continuum absorption (by N$_2$, H$_2$O and CO$_2$) are considered. N$_2$ continuum data is taken from \citet{borysow1986n2n2} and \citet{lafferty1996}, whereas H$_2$O self and foreign continua as well as the CO$_2$ foreign continuum are incorporated following \citet{clough1989}. The CO$_2$ collision-induced self continuum is described following \citet{kasting1984}. In the lower atmosphere, the model performs convective adjustment such that temperature profiles follow the wet adiabat. The formulation of the adiabatic lapse rate takes into account the condensation of H$_2$O or CO$_2$ \citep{kasting1991}. The super-saturation ratio for the onset of CO$_2$ condensation is unity (but see \citet{glandorf2002} for a different estimate for early Mars).

H$_2$O profiles are re-calculated for each model time step, according to local temperature and using a fixed relative humidity (RH) profile. The concentrations of the other species are fixed at the start of the calculations and only adjusted according to condensation of water and carbon dioxide at the surface \citep{vparis2013max}.

For more details, we refer to \citet{vparis2008} and references therein.

\subsection{Model improvements}

\label{improv}

The original IR radiative transfer scheme MRAC (as used in, e.g., \citealp{vparis2008}, \citealp{vparis2010gliese}, \citealp{vparis2013marsn2}) only considered gaseous absorption by H$_2$O and CO$_2$ for the calculation of opacities. For this work, a new, updated MRAC version has been created that also considers CH$_4$ and O$_3$ for gaseous absorption.

Table \ref{intervals} shows the spectral intervals as well as the species included in these intervals.

\begin{table*}
 \caption{Spectral intervals for the new IR radiative transfer scheme and species considered} \centering
\begin{tabular}{ccc}

\hline\hline
 Interval & range [$\rm{cm}^{-1}$] & contributing species\\
 \hline\hline
 1  & 7,470 - 10,000 & CO$_2$,H$_2$O,CH$_4$ \\
 2  & 6,970 - 7,470  &  CO$_2$,H$_2$O,CH$_4$ \\
 3  & 6,000 - 6,970  &  CO$_2$,H$_2$O,CH$_4$\\
 4  & 5,350 - 6,000  &  CO$_2$,H$_2$O,CH$_4$ \\
 5  & 4,600 - 5,350  & CO$_2$,H$_2$O,CH$_4$ \\
 6  & 4,100 - 4,600  & CO$_2$,H$_2$O,CH$_4$ \\
 7  & 3,750 - 4,100  &  CO$_2$,H$_2$O,CH$_4$\\
 8  & 3,390 - 3,750  &  CO$_2$,H$_2$O,CH$_4$ \\
 9  & 3,050 - 3,390  &  CO$_2$,H$_2$O,CH$_4$ \\
 10 & 2,750 - 3,050  & CO$_2$,H$_2$O,CH$_4$ \\
 11 & 2,400 - 2,750  &  CO$_2$,H$_2$O,CH$_4$ \\
 12 & 2,250 - 2,400  & CO$_2$,H$_2$O,CH$_4$ \\
 13 & 2,150 - 2,250  &  CO$_2$,H$_2$O,CH$_4$,O$_3$ \\
 14 & 2,000 - 2,150  &  CO$_2$,H$_2$O,CH$_4$,O$_3$ \\
 15 & 1,850 - 2,000  &  CO$_2$,H$_2$O,CH$_4$,O$_3$ \\
 16 & 1,400 - 1,850  & CO$_2$,H$_2$O,CH$_4$,O$_3$ \\
 17 & 1,100 - 1,400  &  CO$_2$,H$_2$O,CH$_4$,O$_3$ \\
 18 & 1,000 - 1,100  &  CO$_2$,H$_2$O,CH$_4$,O$_3$ \\
 19 & 905 - 1,000    &  CO$_2$,H$_2$O,CH$_4$,O$_3$ \\
 20 & 820 - 905      &  CO$_2$,H$_2$O,CH$_4$,O$_3$ \\
 21 & 730 - 820      &  CO$_2$,H$_2$O,O$_3$ \\
 22 & 600 - 730      &  CO$_2$,H$_2$O,O$_3$ \\
 23 & 525 - 600      &  CO$_2$,H$_2$O,CH$_4$,O$_3$ \\
 24 & 460 - 525      &  CO$_2$,H$_2$O,CH$_4$ \\
 25 & 20 - 460       &  H$_2$O,CH$_4$,O$_3$ \\

\end{tabular}

\label{intervals}

\end{table*}

Line positions and line strengths for both CH$_4$ and O$_3$ have been taken from the Hitran 2008 database \citep{rothman2009}. Line strengths are converted from the reference temperature of 296\,K to the desired temperature following \citet{norton1991}. Cross sections are then obtained with the MIRART-Squirrl line-by-line radiative transfer code \citep{schreier2001,schreier2003} using 10$^6$ spectral points per band. The line shape is taken to be a Voigt line profile with a cutoff of 10\,cm$^{-1}$. The foreign component of the Lorentz broadening is taken directly from the Hitran database, i.e. for air. The self component of the Lorentz broadening was calculated assuming a volume mixing ratio (vmr) of 10$^{-6}$ for both CH$_4$ and O$_3$, 3.55$\cdot$10$^{-4}$ for CO$_2$ and 10$^{-3}$ for water. Numerical tests showed that except for high water concentrations ($\gtrsim$10$^{-2}$), the effect of the self component on the cross sections was negligible.

The chosen line cutoff in this work is relatively short. Tests with a line-by-line radiative transfer code with line cut-offs at 25\,cm$^{-1}$ have shown that the overall flux change rarely exceeds 5\,\% in a given spectral band. For the total integrated outgoing thermal flux, we find a difference of about 1\,\%, equal to a radiative forcing of a few W\,m$^{-2}$. This could lead to a small change in surface temperature of a few K (see also, for example, \citealp{wordsworth2010cont}). However, this is not expected to qualitatively alter our conclusions. Furthermore, as pointed out by, e.g., \citet{halevy2009}, \citet{wordsworth2010cont} or \citet{mischna2012}, assuming a Voigt line profile is probably not a well-justified choice in dense CO$_2$ atmospheres where the far wings of lines are substantially sub-Lorentzian. Sensitivity tests by \citet{wordsworth2010cont} or \citet{vparis2010gliese} showed that the influence on surface temperature can be large (of the order of 10\,K). Still, this is again not likely to influence the conclusions regarding the impact of ozone on the thermal structure in the upper and mid atmosphere.

The interpolation variables are temperature, log(pressure) and the individual concentrations $c_i$ of the absorbing species. The concentration grid has four dimensions, one for H$_2$O, CO$_2$, CH$_4$ and O$_3$, respectively. Table \ref{includedphysics} summarizes the range of the interpolation variables and the type of interpolation.

\begin{figure}[h]
\begin{center}
  \includegraphics[width=240pt]{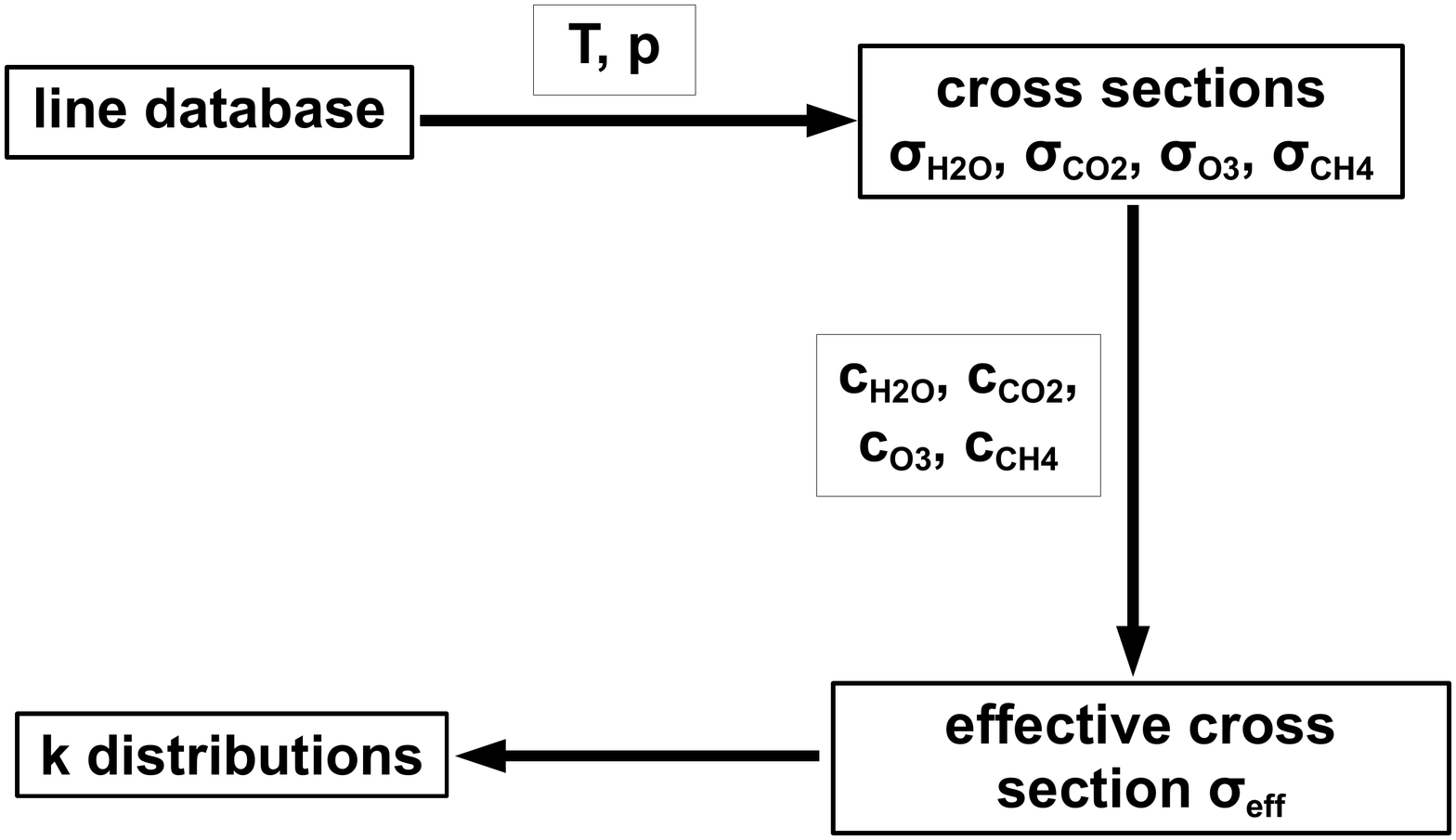} \\
\caption{Flowchart for calculation of k distributions.} 
\label{flow}
\end{center}
\end{figure}

Figure \ref{flow} illustrates our approach to calculate the needed k distributions. In every spectral interval, for each temperature-pressure point (T/p), cross sections $\sigma_i$ have been calculated for the contributing species $i$ (see Table \ref{intervals}). These are then combined to obtain effective cross sections $\sigma_{\rm{eff}}$ for the different gas mixtures according to assumed molecular concentrations $c_i$ from Table \ref{includedphysics}:

 \begin{equation}
\label{effect}
\sigma_{\rm{eff}}=\sum_i c_i \cdot \sigma_i
\end{equation}

From the effective cross sections $\sigma_{\rm{eff}}$, k distributions have been obtained that are then used in the model for the radiative transfer calculations with a correlated-k approach.

\begin{table*}
\centering
\caption[Interpolation and range of the grid
variables]{Interpolation and range of the grid variables in the new
IR radiative transfer scheme}

\begin {tabular}{lccc}

\hline\hline

Quantity    & Range     & Interpolation & Comments \\
\hline \hline
Temperature & 100-400 K & linear in T   & $\Delta$\,T=50 K\\

Pressure    & 10$^{-5}$-1.5 bar & linear in log(p)   & $\Delta$\,log(p)=0.1\\

H$_2$O concentration& 10$^{-9}$-1 & linear in vmr   & 1-4 points per order of magnitude\\

carbon dioxide concentration& 10$^{-6}$-1 & linear in vmr   & 1-4 points per order of magnitude\\

CH$_4$ concentration& 10$^{-8}$-10$^{-2}$ & linear in vmr   & 1-4 points per order of magnitude\\

O$_3$ concentration& 10$^{-8}$-10$^{-2}$ & linear in vmr   & 1-4 points per order of magnitude\\

\end{tabular}\label{includedphysics}
\end{table*}

\subsection{Model verification}

\begin{figure}[h]
\begin{center}
  \includegraphics[width=400pt]{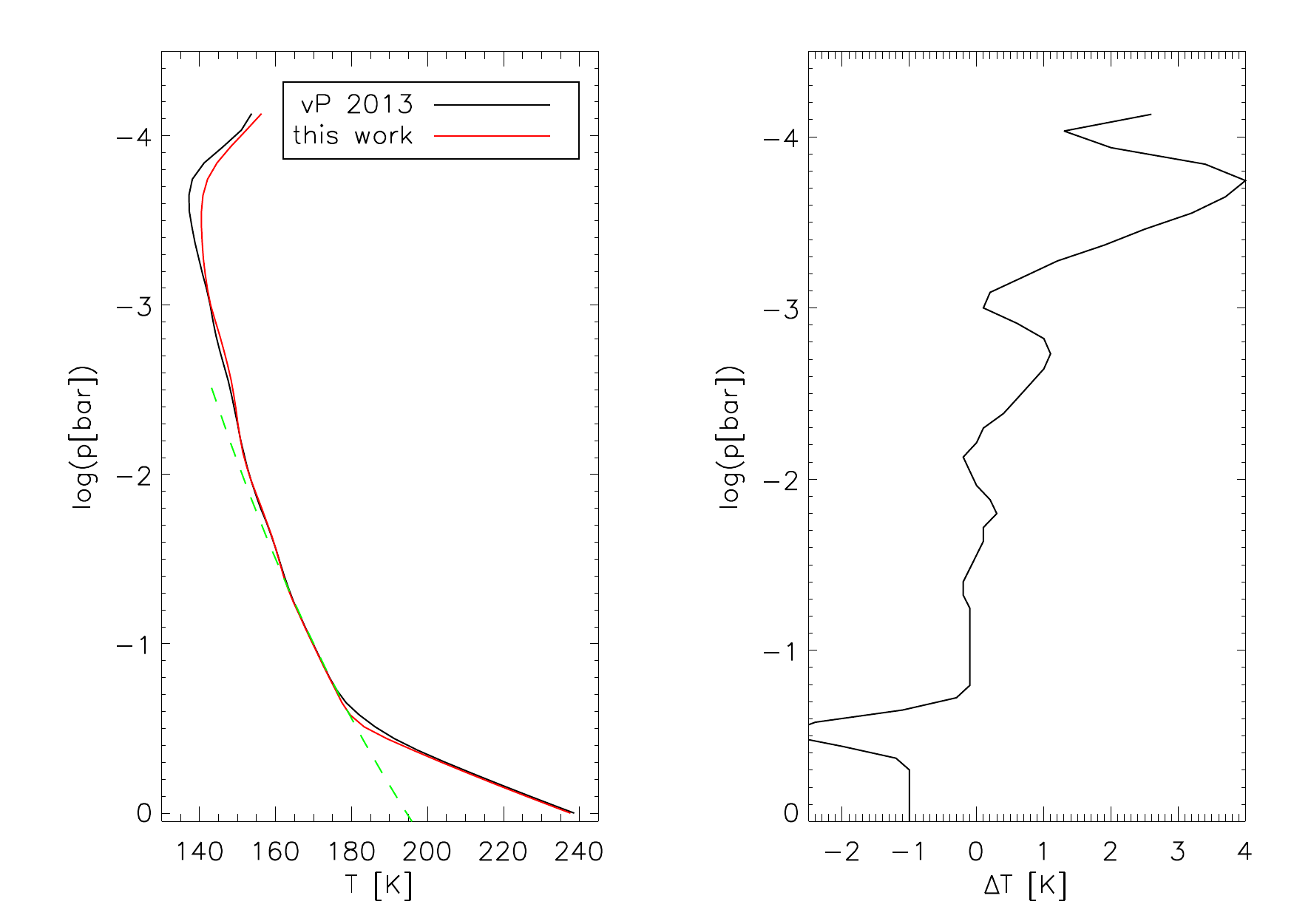} \\
\caption{Early Mars: Comparison of temperature profiles from \citet{vparis2013marsn2} and this work. Carbon dioxide saturation vapor pressure curve as green dashed line.} 
\label{tprofmars}
\end{center}
\end{figure}

We tested the new model against results for early Mars from \citet{vparis2013marsn2}. As an example, Figure \ref{tprofmars} shows the temperature profiles and differences for a 1\,bar carbon dioxide case (without N$_2$). It is clearly seen that both models compare relatively well with each other. The small discrepancies are due to a slightly different approach to calculate the line broadening parameters for the k distributions at high pressures, compared to earlier work (e.g., \citealp{vparis2008,vparis2010gliese}). In the line-by-line radiative transfer code used to calculate the cross sections (see above), the broadening pressure $p_b$ of the gas species i is calculated from the ideal gas law, i.e. $p_b=c_i \cdot p$ instead of using a column-density correction to calculate the actual broadening pressure (e.g., see \citealp{kasting1987}). However, these differences do not affect the conclusions of previous studies.

The second model verification uses a modern Earth reference profile. Ozone and methane concentrations were taken from \citet{grenfell2011}. We run simulations with two different thermal radiative transfer schemes. First, we used RRTM \citep{Mlawer1997},  a scheme that is widely used in 1D and 3D exoplanet and Earth climate studies (e.g., \citealp{Seg2003,Seg2005}, \citealp{roeckner2006}, \citealp{grenf2007pss}, \citealp{kaltenegger2011}, \citealp{stevens2013}). A second simulation was done with the new scheme presented in this work.

\begin{figure}[h]
\begin{center}
  \includegraphics[width=400pt]{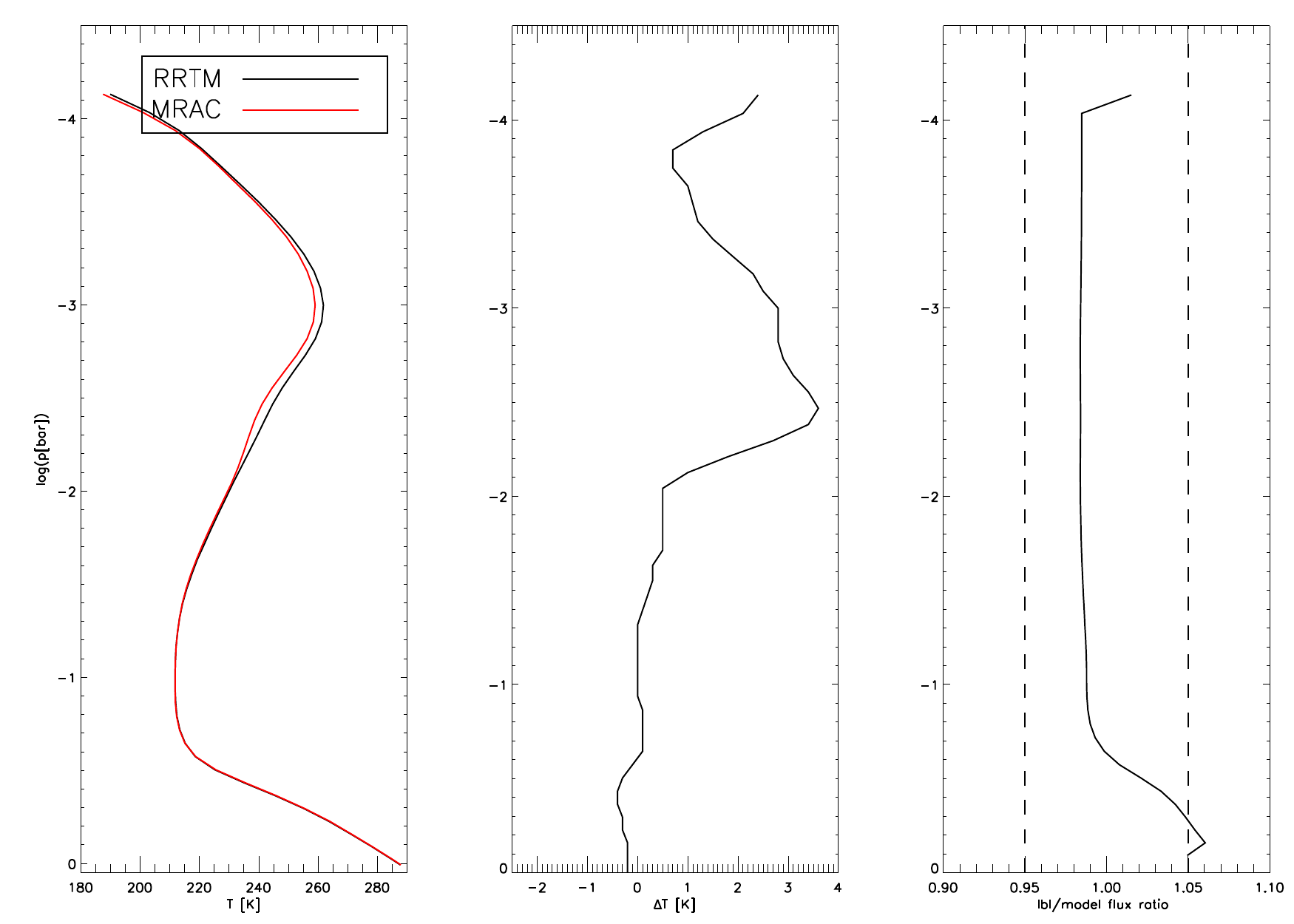}  \\
\caption{Modern Earth model verification: (Left panel) Comparison of temperature profiles calculated with RRTM \citep{Mlawer1997} and MRAC (this work). (Center panel) $\Delta$T (RRTM-MRAC) between both models, (Right panel) IR upwards flux ratio (climate model)/line-by-line.}
\label{tprofb}
\end{center}
\end{figure}

Figure \ref{tprofb} shows the resulting temperature profiles (left panel), the temperature difference between both models (center panel) and a flux comparison for the upwelling thermal flux (right panel). For the latter, we compared calculated model IR fluxes with fluxes calculated by the high-resolution line-by-line code MIRART-Squirrl.

Concerning the temperature profiles, the agreement in the troposphere and the lower stratosphere is very good. In the upper stratosphere, small discrepancies can be seen, reaching up to about 3\,K. However, the overall agreement is good, and such small temperature differences are not expected to have a large influence neither on observables (e.g. spectral signatures) nor on temperature-dependent chemistry. For the IR thermal upwelling flux, it is clearly seen that the fluxes do not differ by more than a few \% throughout the stratosphere.

Based on these verifications, we consider the new radiative scheme to be applicable to further planetary scenarios.

\subsection{Early Mars scenarios}

Simulated model atmospheres were assumed to be composed of CO$_2$, H$_2$O and O$_3$. The CO$_2$ partial pressure is fixed at 1\,bar, consistent with upper limits placed on atmospheric pressure during the Noachian (e.g., \citealp{phillips2001}). H$_2$O is calculated according to ambient temperature, assuming a fully saturated atmosphere (i.e., RH=1, see \citealp{vparis2013marsn2}). This approach is similar to most 1D studies of early Mars (e.g., \citealp{mischna2000}, \citealp{Cola2003}, \citealp{tian2010}), but most likely over-estimates the amount of atmospheric water. The surface albedo was set to 0.21, close to the observed value of present Mars \citep{kieffer1977}. This value of surface albedo allows the model to reproduce present Mars mean surface temperatures. Planetary gravity is 3.73\,ms$^{-2}$, and the orbital distance is set to 1.52\,AU. We neglected any effect of eccentricity. However, even though Mars' orbit is eccentric today ($e$=0.09), and eccentricity cycles would occasionally drive the eccentricity to high values (e.g., \citealp{laskar2004}), the overall effect would probably be small (at $e$=0.2, the mean flux increases by only 2\,\% compared to the circular case). 

The incoming solar irradiation was set to Noachian conditions 3.8 billion years ago, i.e. 75\,\% of today's irradiation (e.g., \citealp{gough1981}). The input spectrum was taken from \citet{gueymard2004} and scaled at all wavelengths accordingly.  Note that this approach is consistent with previous early Mars studies (e.g., \citealp{kasting1991}, \citealp{tian2010}, \citealp{wordsworth2013}). Other work, primarily on early Earth climate, not only scaled the spectrum, but also incorporate the variation of stellar parameters (e.g., \citealp{goldblatt2009faintyoungsun}) and enhanced UV radiation (e.g., \citealp{kunze2014}). Enhanced UV radiation (see, e.g., \citealp{ribas2005}) of the young Sun could have played an especially important role affecting ozone hence the thermal structure of the atmosphere.

Ozone profiles strongly depend on the assumed oxygen content of the atmosphere, the incoming stellar UV radiation, and, in addition, on the catalytic cycles (HO$_x$, NO$_x$, etc.) operating in the atmosphere (e.g., \citealp{selsis2002}, \citealp{segura2007}, \citealp{grenfell2013se_chem}). Detailed photochemical modeling is warranted for this problem, however such relatively complex models would necessitate many more unconstrained boundary conditions and parameters (such as surface fluxes of oxidizing and reducing compounds, wet and dry deposition rates, vertical mixing, kinetic data, UV radiation field, etc.). Such complex models (e.g., \citealp{selsis2002}, \citealp{segura2007}, \citealp{hu2012chem}, \citealp{domagal2014}, \citealp{tian2014}, \citealp{grenfell2014}) allow for an estimate of possible ranges of ozone concentrations. Motivated by the results from detailed photochemical studies, we perform a sensitivity study of the influence of ozone on temperature structure by inserting artificial ozone profiles $C_{O_3}$. These profiles are not the output of a photochemical model, but are parameterized as a function of pressure $p$ based on a Gaussian profile

\begin{equation}\label{o3profile_para}
 C_{O_3}\left(p\right)=\rm{vmr}_{\rm{max}}\cdot exp\left(-\frac{\left(log\frac{p}{p_{\rm{max}}}\right)^2}{0.5}\right)
\end{equation}

where $p_{\rm{max}}$ is the pressure of the maximum concentration and $\rm{vmr}_{\rm{max}}$ is the maximum volume mixing ratio. 

\begin{figure}[h]
\centering
  \includegraphics[width=300pt]{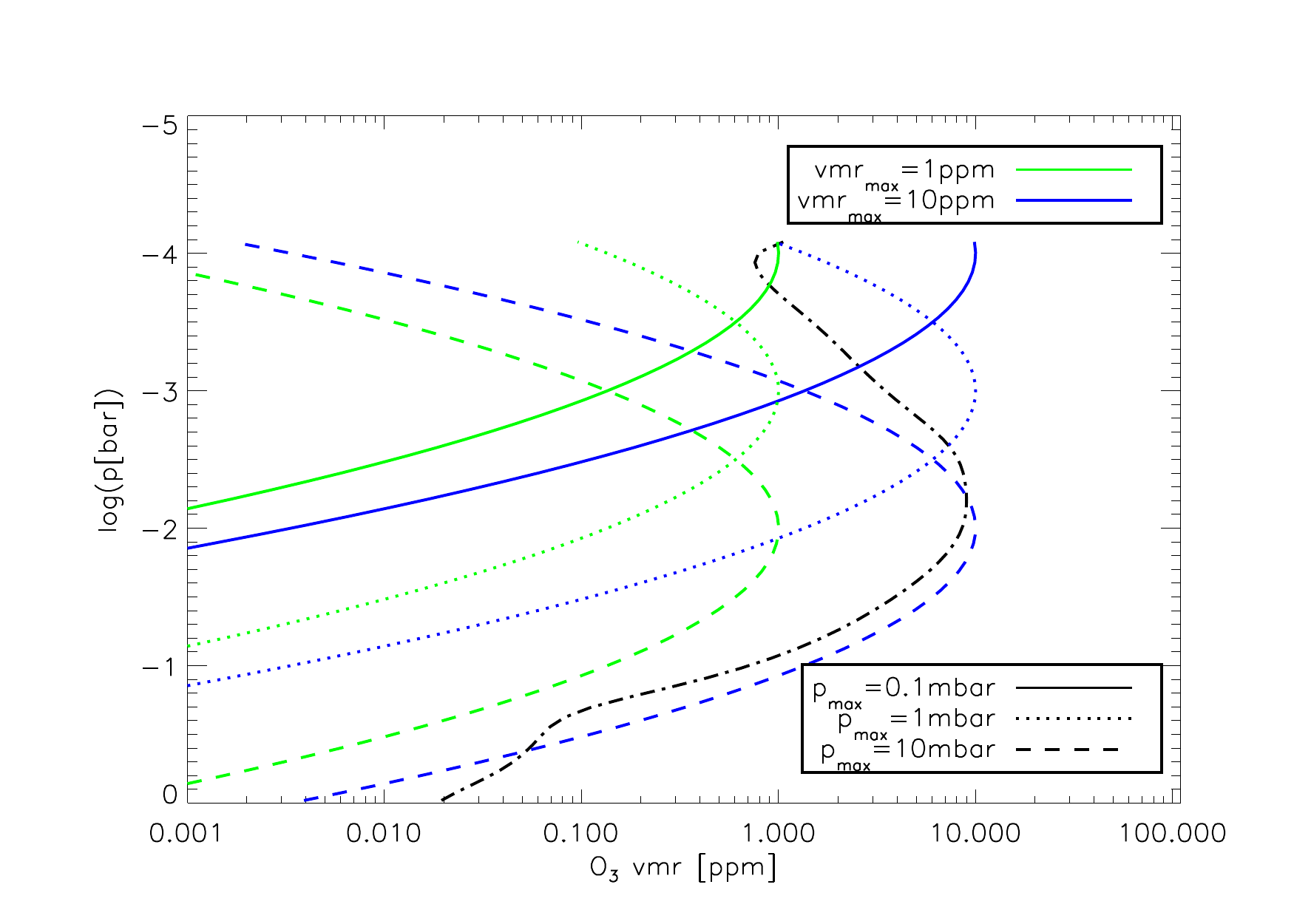} \\
\caption{Sample of chosen ozone profiles. Earth mean profile from \citet{grenfell2011} as dash-dotted line.}
\label{o3inputprof}
\end{figure}

We performed simulations for different combinations of $\rm{vmr}_{\rm{max}}$ and $p_{\rm{max}}$, as shown in Table \ref{tableo3}. In Fig. \ref{o3inputprof}, we show a sample of the chosen ozone profiles, together with a modern Earth mean profile.

\begin{table}[h]
\centering
\caption{Adopted parameter values for eq. \ref{o3profile_para} in this work.}

\begin {tabular}{lc}

\hline\hline

Parameter   & Values \\
\hline \hline
$p_{\rm{max}}$ [mbar] & 0.1, 1, 10\\

$\rm{vmr}_{\rm{max}}$  [ppm]    & 0.1, 1, 10, 100\\

\end{tabular}\label{tableo3}
\end{table}

The ozone layer peak (its location and its magnitude) is the result of an interplay between two main effects. On the one hand, ozone formation requires photolytic release ($\lambda \lesssim 200$\,nm) of atomic oxygen (O) e.g. from O$_2$, CO$_2$, H$_2$O, which are usually more abundant on the lower levels (followed by fast, three-body formation of ozone). On the other hand, UV radiation is more abundant with increasing altitude, hence there is generally a distinct maximum of ozone concentration at pressures of about 10$^{-4}$-10$^{-2}$\,bar. This motivates the form of our artificial ozone profiles.


Previous calculations by, e.g., \citet{selsis2002} and \citet{segura2007} suggest a maximum in ozone number density of up to 10$^{12}$\,cm$^{-3}$, located at pressures around 1-10\,mbar. They found ozone columns between 10$^{-4}$-2 times the modern Earth column (Table 3 in \citealp{selsis2002} and Table 2 in \citealp{segura2007}). Both studies focused on N$_2$- or CO$_2$-rich atmospheres, similar to the ones considered here. The oxygen needed for ozone formation was either provided by H$_2$O or CO$_2$. To cover the entire range of possible ozone concentrations found by \citet{selsis2002} (i.e., their humid and dry CO$_2$ cases as possible extremes) we varied the maximum ozone concentration between 0.1 and 100\,ppm. As shown in Table \ref{columno3}, this leads to relatively thick ozone columns (up to 24x the modern Earth values) for a few cases.

\begin{table}[h]
\centering
\caption{O$_3$ columns (in units of terrestrial columns, where one terrestrial column corresponds to 315 Dobson units) for the scenarios listed in Table \ref{tableo3}.}

\begin {tabular}{|l|cccc|}

\hline

\backslashbox{$p_{\rm{max}}$ [mbar] }{$\rm{vmr}_{\rm{max}}$  [ppm]    }  & 0.1 &1 & 10& 100 \\
\hline 
0.1& 0.00023  &  0.0023  &   0.023    & 0.23 \\
1  &  0.00249   &  0.0249  &  0.249    &   2.48 \\
10  & 0.0248 &  0.248    &   2.47     &  24.7\\
\hline

\end{tabular}\label{columno3}
\end{table}

\section{Results and Discussion}

\label{results}

\subsection{Surface temperature}

\begin{figure}[h]
\centering
  \includegraphics[width=400pt]{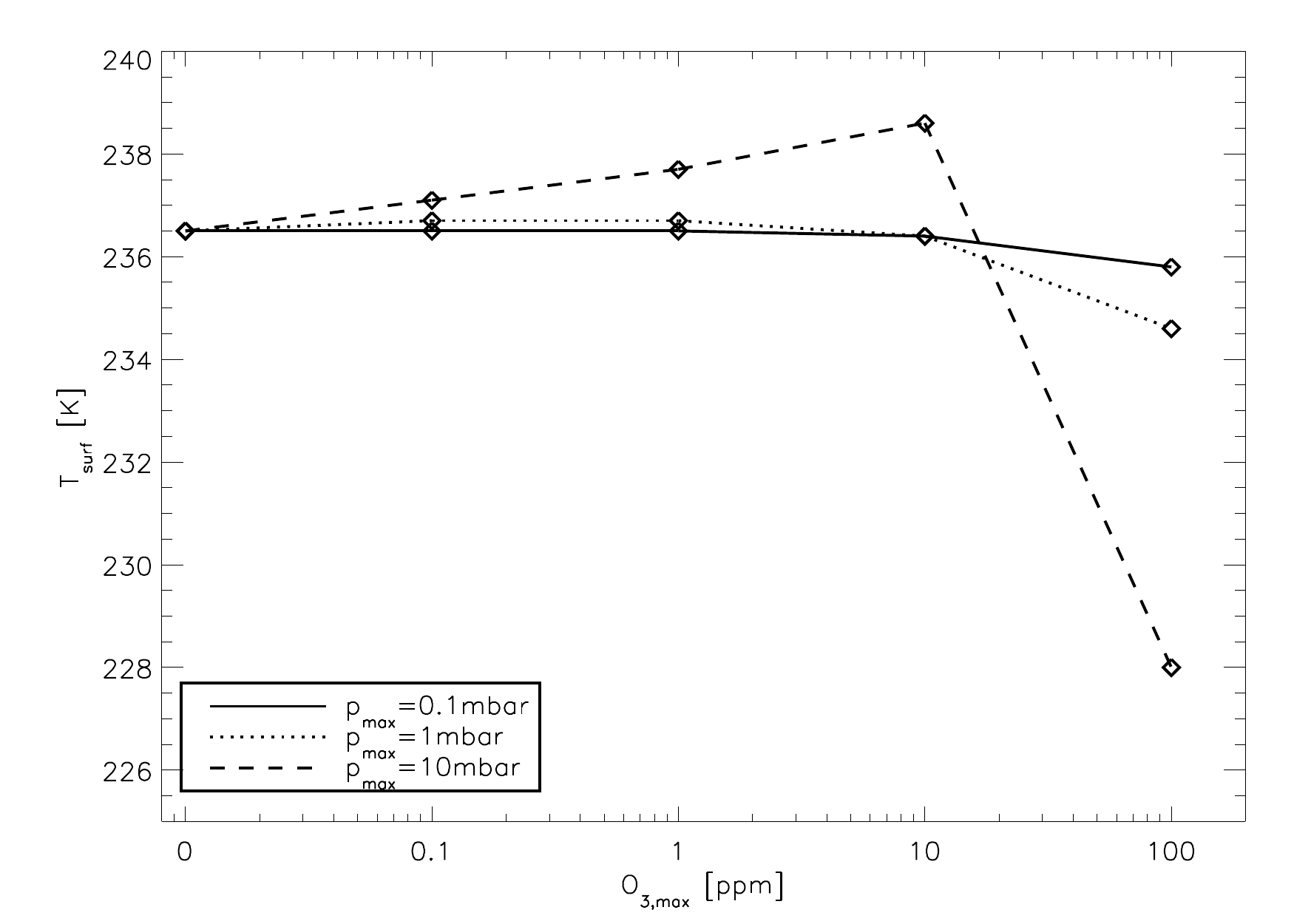} \\
\caption{Effect of ozone on surface temperature.}
\label{marso3tsurf}
\end{figure}

Figure \ref{marso3tsurf} shows the surface temperature as a function of the parameters in eq. \ref{o3profile_para}.

For all scenarios, the surface temperatures first slightly increase with increasing $\rm{vmr}_{\rm{max}}$, due to a slight decrease of planetary albedo and an additional greenhouse effect from ozone. For higher $\rm{vmr}_{\rm{max}}$, surface temperatures start to decrease with $\rm{vmr}_{\rm{max}}$ (see below).

However, for the $p_{\rm{max}}$=0.1 and 1\,mbar cases, the overall effect is rather small, of the order of 1-2\,K. In contrast, for $p_{\rm{max}}$=10\,mbar, the increase at low $\rm{vmr}_{\rm{max}}$ (up to 2\,K) and the subsequent decrease towards higher values is much more pronounced, with a maximum temperature drop of about 8\,K compared to the zero-ozone case.

\subsection{Temperature structure}

\begin{figure}[h]
\centering
  \includegraphics[width=400pt]{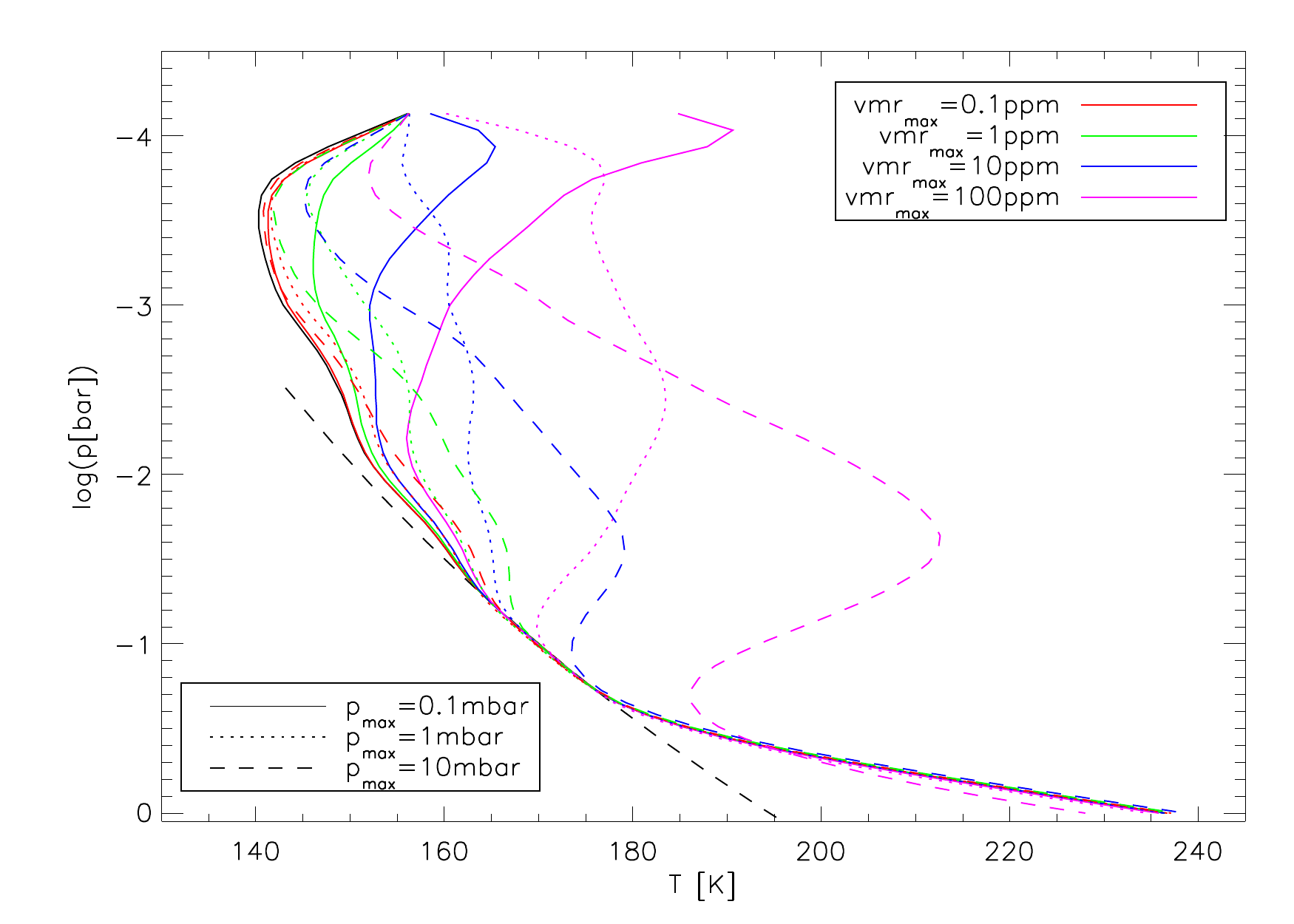} \\
\caption{Effect of ozone on temperature structure. Reference case (zero ozone) as black plain line. Carbon dioxide saturation pressure shown as black dashed line.}
\label{marso3temp}
\end{figure}

Figure \ref{marso3temp} shows the effect of ozone on the calculated temperature-pressure profiles. In the middle and upper atmosphere, up to 60\,K of warming is seen, depending on the choice of O$_3$ parameters.  This is of course due to the massive increase in stellar heating rates associated with the absorption of solar UV absorption by ozone. Heating rates increase by up to a factor of 5 compared to the no-ozone case (not shown).

Two other effects are readily inferred from Fig. \ref{marso3temp}. First, for the $p_{\rm{max}}$=10\,mbar cases at $\rm{vmr}_{\rm{max}}$=10 and 100\,ppm, temperature profiles no longer follow the carbon dioxide saturation pressure curve (black dashed line in Fig. \ref{marso3temp}). This suggests that carbon dioxide condensation, hence also the formation of carbon dioxide clouds, is prohibited throughout the entire atmosphere. For the other cases, the potential cloud formation zone is reduced, although the effect is rather minimal at low $\rm{vmr}_{\rm{max}}$ or small $p_{\rm{max}}$. Second, as suggested by the slope of the $\rm{vmr}_{\rm{max}}$=100\,ppm,  $p_{\rm{max}}$=10\,mbar case, the lower atmosphere is no longer convective. This is in contrast to the other scenarios, where the lapse rate follows a CO$_2$ adiabat in the mid atmosphere and a wet H$_2$O adiabat in the lower atmosphere.

\begin{figure}[h]
\centering
  \includegraphics[width=400pt]{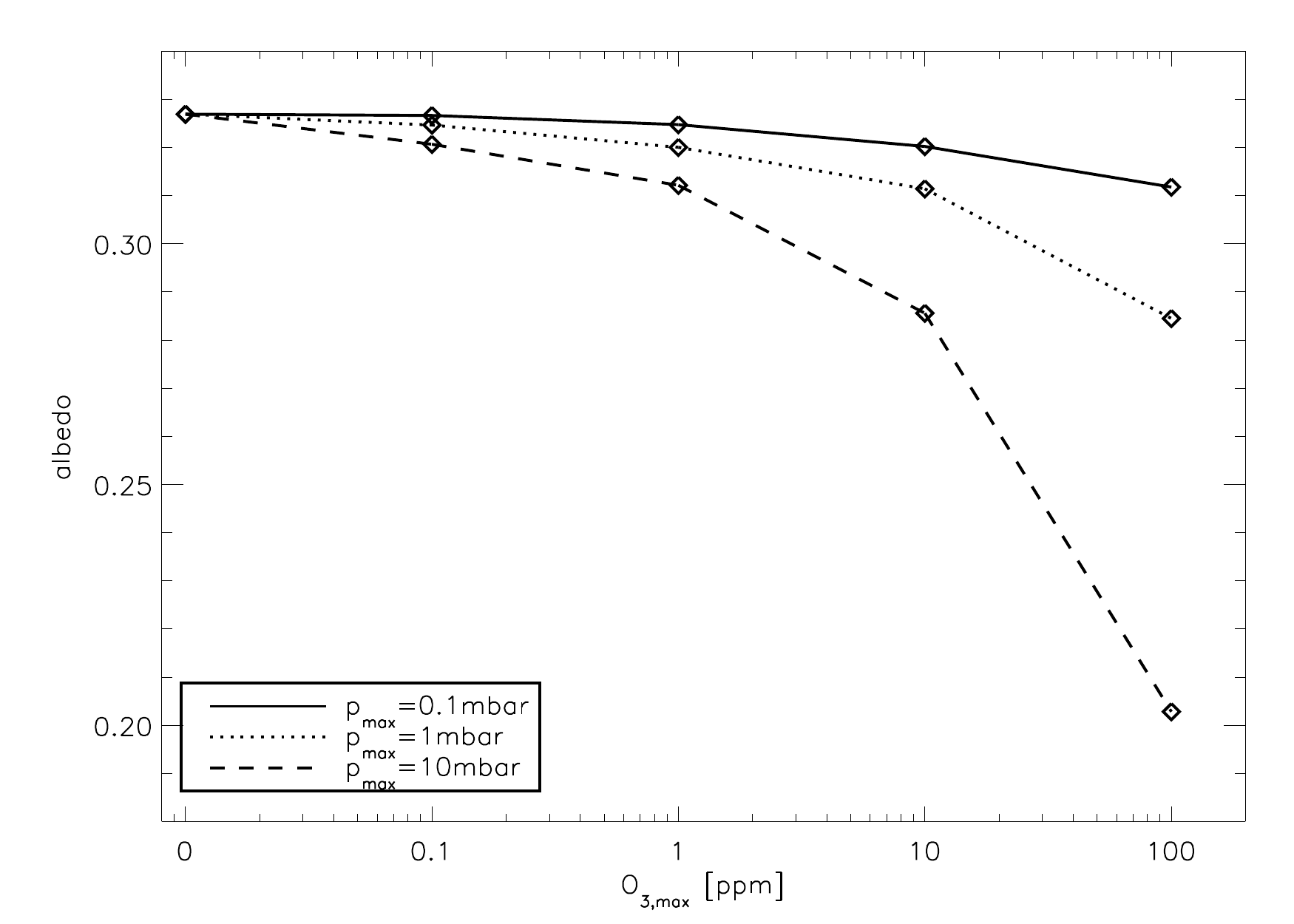} \\
\caption{Effect of ozone on calculated planetary albedo.}
\label{marso3albedo}
\end{figure}

\begin{figure}[h]
\centering
  \includegraphics[width=400pt]{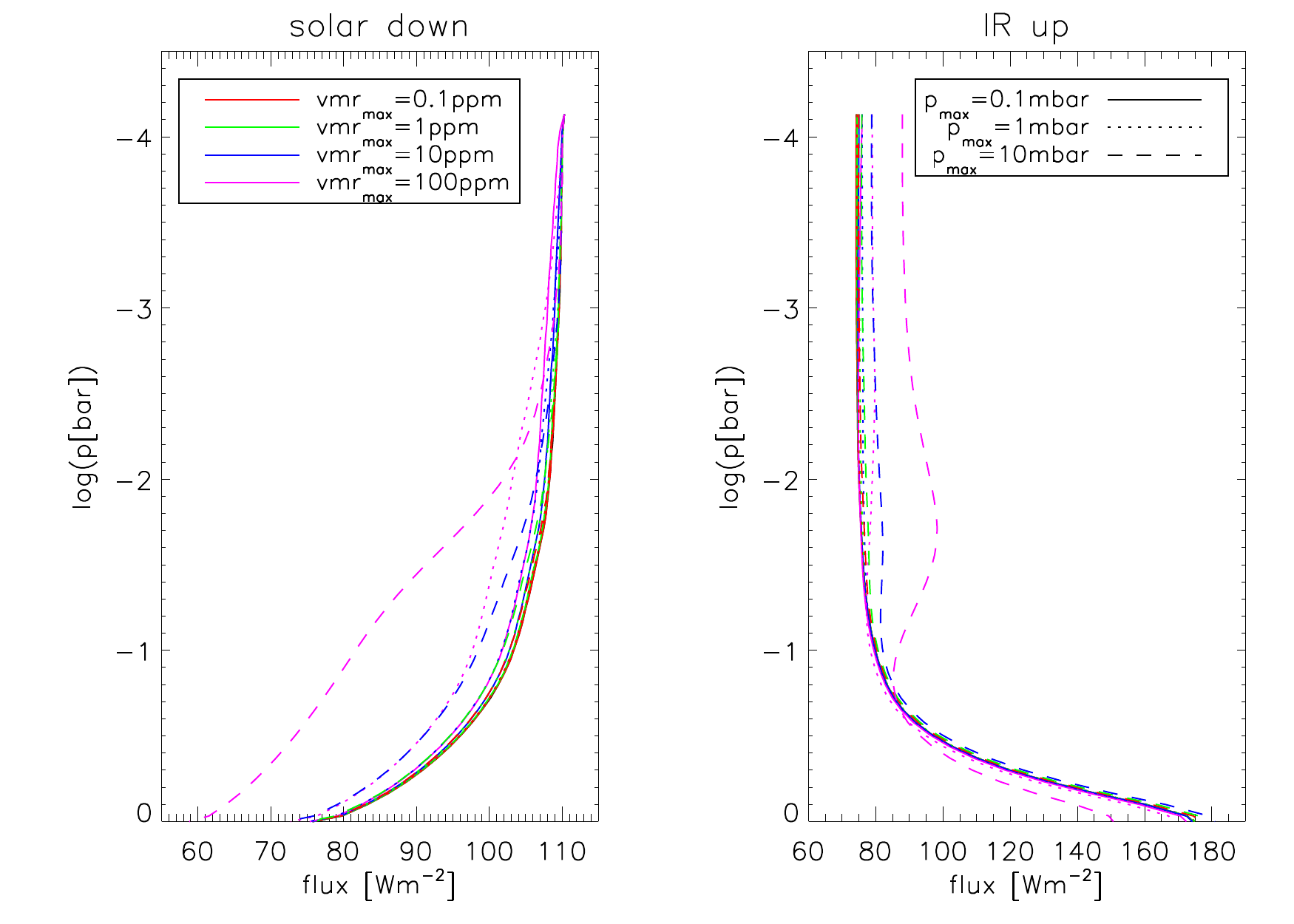}\\
\caption{Effect of ozone on downwards solar (left) and upwelling thermal fluxes (right). Captions correspond to both panels.}
\label{marso3ir}
\end{figure}

The strong UV absorption by ozone drastically changes the energy balance in the atmosphere. Figure \ref{marso3albedo} illustrates the effect of ozone on the planetary albedo. The planetary albedo is reduced by up to 40\,\% at large ozone layers. In terms of additional radiative energy deposited into the atmosphere, a reduction of albedo from 0.32 (zero-ozone) to 0.2 (maximum ozone effect) corresponds to a forcing of 13\,Wm$^{-2}$. However, as illustrated in the left panel of Fig. \ref{marso3ir}, most of this energy is absorbed in the middle atmosphere. As a consequence of the strong absorption, the incoming solar flux at the surface is reduced by about 20\,\%. Therefore, the total radiative flux in the lower atmosphere (sum of stellar and thermal fluxes) is reduced, coming closer to radiative equilibrium. Hence, less energy is available to drive convection. This then leads to a radiative instead of a convective lower atmosphere (as observed in Fig. \ref{marso3temp}).

A reduced albedo corresponds to a larger thermal flux that must be emitted by the atmosphere to satisfy energy balance and radiative equilibrium at the top of the atmosphere. For optically thin atmospheres such as Earth's, this generally leads to an increase in surface temperature. In the case of dense carbon dioxide-dominated atmospheres considered here, the thermal emission originates mostly from pressures at around 10-100\,mbar, since at higher pressures, the atmosphere is optically thick for thermal radiation at all wavelengths (see right panel of Fig. \ref{marso3ir}). For the considered scenarios with a substantial increase of local temperature (see Fig. \ref{marso3temp}), the local IR flux also increases. This balances the solar flux deposited in the atmosphere. Most of the outgoing thermal flux stems from emission in the 15\,$\mu$m band of carbon dioxide which becomes optically thick at pressures around 10-100\,mbar.

\subsection{Discussion}

\subsubsection{Oxygen content and catalytic chemistry}

The assumed ozone profiles are dependent on the oxygen content of the early Mars atmosphere. Given that modern Mars has only little oxygen in the atmosphere (around 10$^{-3}$ vmr, e.g., \citealp{yung1999}), it seems likely that our high-ozone cases will over-estimate the total ozone column by a generous margin. However, the formation of ozone also depends on the assumed UV radiation field (e.g., \citealp{segura2007}, \citealp{domagal2010}, \citealp{domagal2014}, \citealp{grenfell2014}, \citealp{tian2014}) which is relatively poorly known for the young Sun \citep{ribas2005}. Observations of younger solar-type stars suggest that far-UV and X-ray emissions strongly decrease with age (e.g., \citealp{ribas2005}) whereas the total luminosity is generally assumed to increase with age (e.g., \citealp{gough1981}, \citealp{caldeira1992}). Therefore, the ratio between oxygen (hence ozone) production via photolysis (in the far-UV) and photolytic ozone destruction (in the near-UV) could presumably change dramatically and produce largely varying ozone concentrations (see, e.g., sensitivity studies by \citealp{segura2007}, \citealp{grenfell2014} or \citealp{tian2014}). Therefore, reliable quantitative estimates of ozone on early Mars are challenging to obtain. Detailed photochemical models would probably only allow for an order-of-magnitude estimate of such quantities, since boundary conditions (such as wet deposition of hydrogenated species and oxidation of reduced volcanic gases) are essentially unconstrained.

In addition, ozone chemistry heavily depends on the operating catalytic cycles in the atmosphere (e.g., \citealp{crutzen1970}, \citealp{grenfell2013se_chem}). On Earth, most of the cycles are related to NO$_x$ and HO$_x$. Possible NO$_x$ sources on early Mars would be cosmic rays (e.g., \citealp{jackman1980}) or lightning (e.g., \citealp{segnav2005}), but these are hard to estimate. It is unclear whether significant HO$_x$ cycles could operate on early Mars, since calculated stratospheres are very dry (see Fig. \ref{watprof}), with vmr about 10$^{-8}$-10$^{-6}$. This is up to several hundred times drier than in the modern Earth stratosphere (H$_2$O vmr of a few times 10$^{-6}$ up to 10$^{-5}$). Note, however, that HO$_x$ cycles are responsible for the low oxygen content in the present Mars atmosphere (e.g., \citealp{nair1994}, \citealp{yung1999}, \citealp{stock2012ica,stock2012pss}) since they quickly recycle the products of CO$_2$ photolysis (CO+O) back to CO$_2$. Without such recycling, the martian atmosphere would be more O$_2$-rich than today, probably at around 5-10\,\% O$_2$. CO$_2$ would still remain the dominant atmospheric species (e.g., \citealp{nair1994}). This can be explained by the fact that at increasingly higher O$_2$ concentrations, O$_2$ UV absorption starts to shield the CO$_2$ from becoming photolyzed since O$_2$ photodissociation cross sections are much larger than corresponding CO$_2$ cross sections (see, e.g., \citealp{yung1999}, \citealp{selsis2002}). Even on Venus with a very dry stratosphere, HO$_x$ cycles operate, using HCl photolysis as source for H \citep{yung1999}. As an alternative destruction cycle, SO$_x$ cycles could be possible, since higher amounts of sulphur-bearing species are expected in the early Mars atmosphere (e.g., \citealp{farq2000}, \citealp{halevy2007}, \citealp{johnson2008}, \citealp{ramirez2014}). 

Investigating these possibilities is however beyond the scope of this sensitivity study that focuses on climatic effects.

\subsubsection{Eddy diffusion and convection}

As a result of the suppression of convection in the lower atmosphere (see Fig. \ref{marso3temp}) for the high-ozone case, the atmosphere becomes stably stratified and vertical transport is significantly reduced. This has a potentially large impact on (photo)chemistry in such atmospheres. In general, photochemical networks include photochemical reactions (including e.g. 3-body reactions, photolysis, decomposition reactions, etc.) as well as a vertical mixing, which is parameterized by so-called Eddy diffusion. For the convective troposphere, 1D photochemical models of Mars or Earth use constant, relatively high, Eddy diffusion coefficients (e.g., \citealp{massie1981}, \citealp{yung1999}). In a stably stratified atmosphere, the diffusion is most likely less efficient. The magnitude of the effect on convection and transport depends, among others, on the topography. Thus, full 3D dynamic simulations are probably needed to investigate the influence on chemistry. This is, for modern Earth and Earth-like \textbf{(terrestrial)} exoplanets, mostly important for biogenic trace gases such as methane and nitrous oxide. In the lower atmosphere, these are more strongly influenced by transport rather than chemistry (e.g., \citealp{Seg2003,Seg2005} \citealp{grenf2007asbio,grenf2007pss}). 

Exploring this issue, while beyond the scope of this work, would however be interesting in the context of fully coupled climate-chemistry simulations.

\subsubsection{Water loss from early Mars}

The very thin atmosphere of current Mars is likely shaped by significant atmospheric loss. Isotopic signatures of atmospheric escape have been found for nitrogen and oxygen (e.g., \citealp{jak2001}, \citealp{fox2010}, \citealp{gillmann2011}). Recent D/H isotopic ratio measurements of the Curiosity rover on Mars suggest that hydrogen escape and water loss has been ongoing over the last few billion years (e.g., \citealp{mahaffy2015}) after the (presumed to be dense) Noachian atmosphere has been eroded. Modeling the escape of water is a very important issue for atmospheric evolution and habitability (e.g., \citealp{kasting1988}, \citealp{kulikov2006}). The water loss rate depends on the stratospheric water content, where water is photolyzed and subsequent loss of H atoms occurs (e.g., \citealp{kasting1988}, \citealp{kasting1993}).

\begin{figure}[h]
\centering
  \includegraphics[width=300pt]{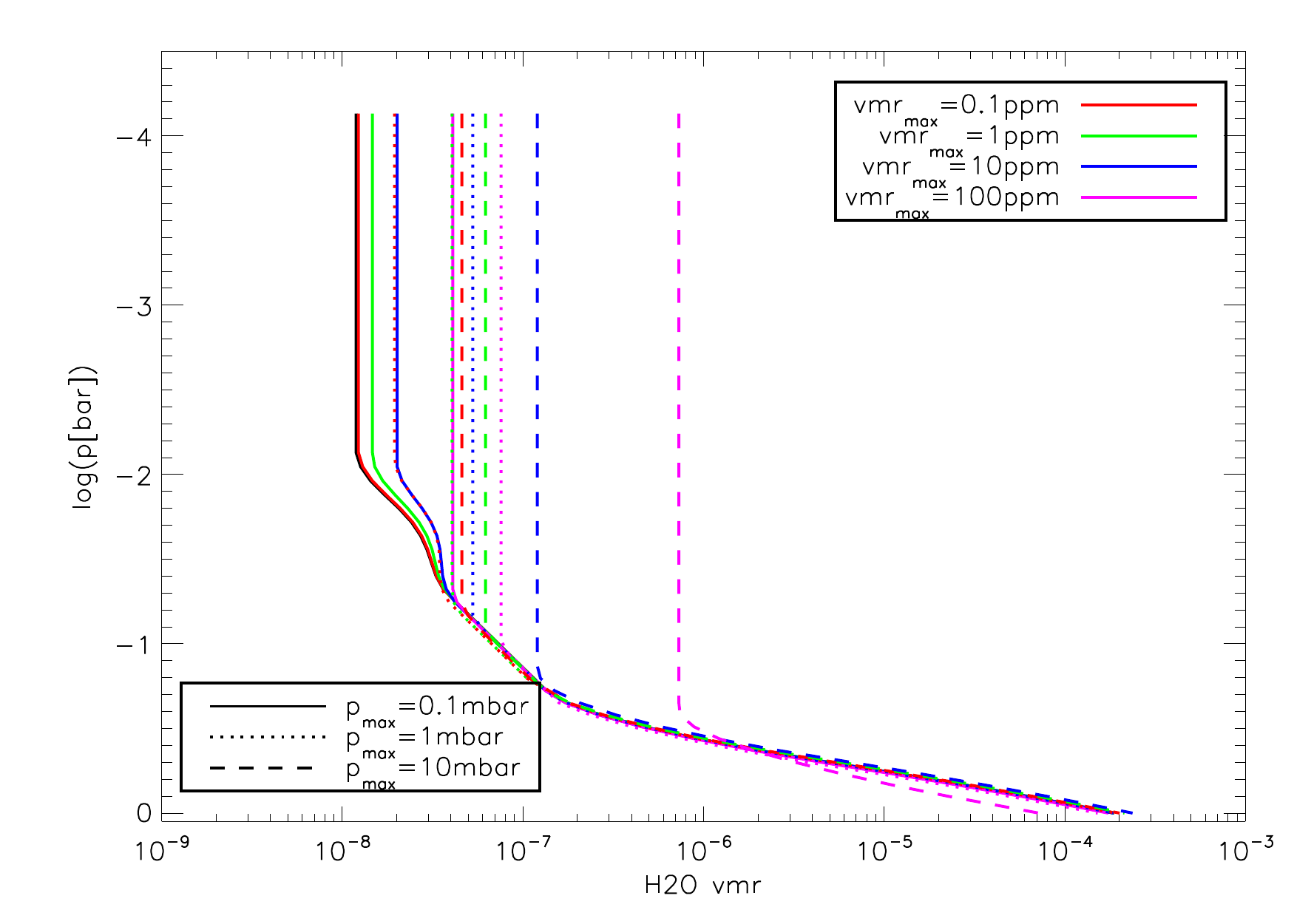}\\
\caption{Effect of ozone on calculated water profiles.}
\label{watprof}
\end{figure}

Figure \ref{watprof} shows the calculated H$_2$O profiles. Note that we assume a relative humidity of unity throughout the atmosphere hence H$_2$O concentrations are likely to be over-estimated in the mid- to upper atmosphere. 
On Earth, the stratospheric H$_2$O content is mainly controlled by the efficient cold trap near the tropopause (around 10\,km altitude). With increasing temperatures due to the ozone layer, water is efficiently trapped in the troposphere. This leads to the dry stratosphere and, consequently, low water escape rates. 

For the early Mars scenarios considered here, cold traps develop even in the absence of temperature inversions. However, given that surface temperatures are very low, and the cold trap is located high in the atmosphere, stratospheric H$_2$O concentrations are very low. This suggests that water loss might be very slow on early Mars. However, when increasing the ozone concentrations in the model atmospheres, the altitude of the cold trap changes dramatically, due to the change in the temperature structure (see Fig. \ref{marso3temp}). This leads to a significant increase of stratospheric H$_2$O content (up to two orders of magnitude), and possibly impacts the water escape rates. Hence, results suggest that water escape rates might be sensitively influenced by the ozone content. Note, however, that an increase in stratospheric H$_2$O concentrations could also lead to enhanced destruction of ozone through catalytic HO$_x$ chemistry (see, e.g., the parametric simulations of \citealp{tian2014}). Therefore, changes in cold-trap locations might limit the increase of ozone, hence limit also H$_2$O escape. To further investigate this issue would need consistent climate-chemistry models to assess this feedback between vertical temperature structure, stratospheric water content and ozone chemistry. 

To estimate the amount of water lost via atmospheric escape, we adopted the same approach as \citet{wordsworth2014}. Escape was assumed to be diffusion-limited and controlled by the cold-trap temperature $T_{\rm{ct}}$ and cold-trap water concentrations. The diffusion coefficient of H$_2$O through a CO$_2$ atmosphere was taken from approximative equations of \citet{marrero1972}. 

These approximations lead to water escape fluxes of, even under the most favorable escape conditions, only about 1\,cm of water lost within 10$^9$ years (1\,Gyr). This has to be compared to the actual atmospheric water column of about a few 10$^{-4}$\,m of water in our simulations.  Furthermore, the original Martian water inventory is estimated to be much larger, about 10$^2$-10$^3$\,m (e.g., \citealp{mckay1989mars}, \citealp{lammer2013}, \citealp{lasue2013}), Currently, most of the water is thought to either reside in the cryosphere or to be bound to minerals in the crust (see discussion in, e.g., \citealp{lammer2013} and \citealp{lasue2013}).

In addition, we point out that current best estimates of the hydrogen escape flux suggest about 1-2\,m of water lost in 1\,Gyr (e.g., \citealp{chass2004}, \citealp{zahnle2015}), about 2 orders of magnitude more than we calculate. This is due to the fact that molecular and atomic hydrogen are the main hydrogen-bearing species in the upper Martian atmosphere, not H$_2$O, as we assumed above. H and H$_2$ have concentrations about 2 orders of magnitude higher than H$_2$O (e.g., \citealp{nair1994}, \citealp{yung1999}). H and H$_2$ are derived from H$_2$O photolysis, and since water is controlled by surface conditions and constantly replenished, H and H$_2$ can build up until concentrations reach an equilibrium between the photolytic source and atmospheric escape. Addressing the connection between cold-trap location, thermal structure and H$_2$O escape necessitates coupled climate-chemistry simulations, which is a potential direction of future research.


\subsubsection{Spectral appearance}

In an exoplanetary context, an interesting question with respect to ozone is the possibility of detecting it remotely with spectroscopic observations. It has been proposed as a so-called biosignature gas (e.g., \citealp{leger1996}, \citealp{selsis2000}, \citealp{schindler2000}, \citealp{DesMarais2002}, \citealp{selsis2002}), and much work has been invested into atmospheric and spectral modeling regarding detection and interpretation of ozone signatures (e.g., \citealp{selsis2002}, \citealp{Seg2003,segura2007}, \citealp{kaltenegger2007}, \citealp{Kaltenegger2009}, \citealp{rauer2011}, \citealp{hedelt2013}, \citealp{grenfell2014}).

\begin{figure}[h]
\centering
  \includegraphics[width=300pt]{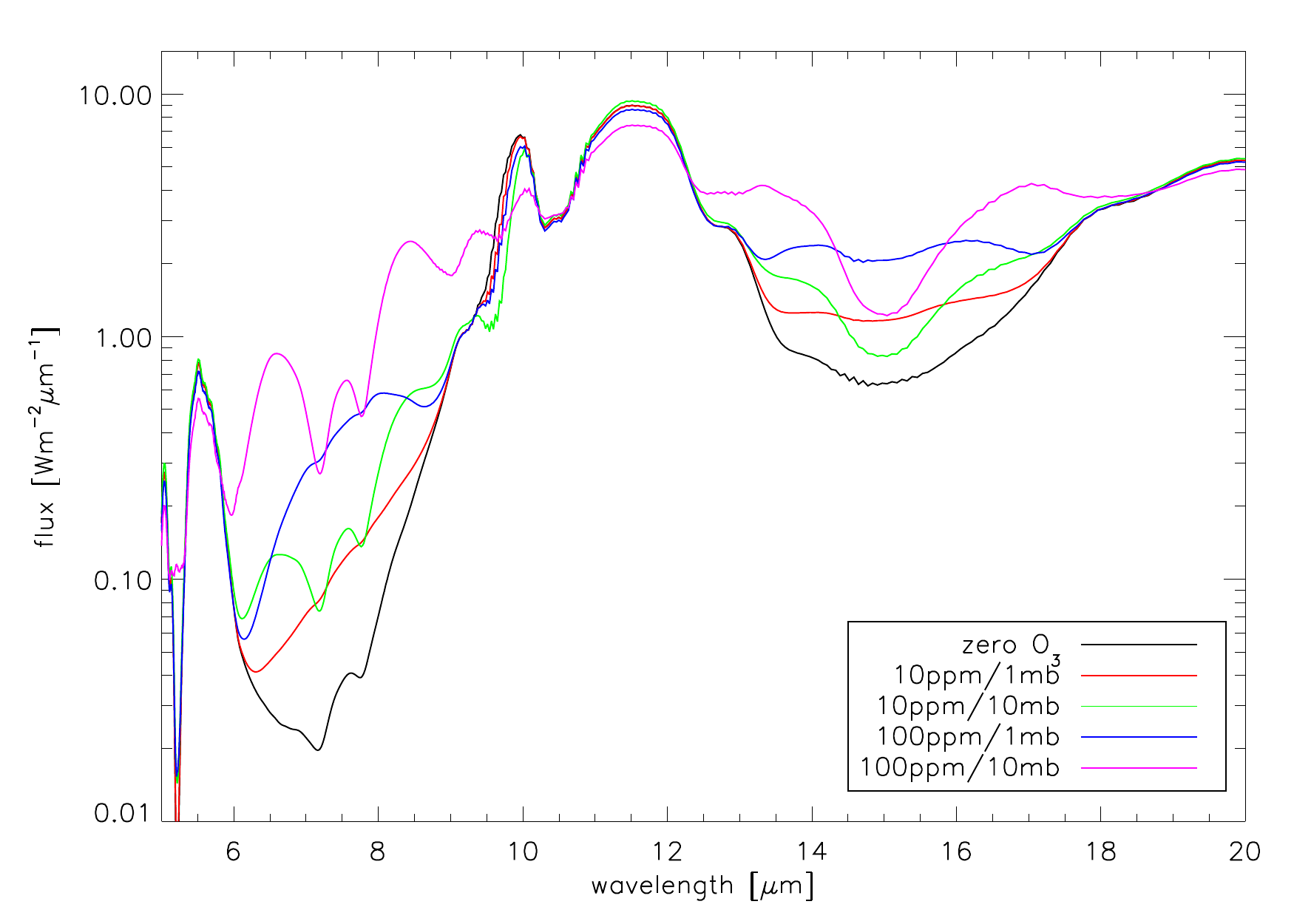}\\
\caption{Emission spectra (spectral resolution R=100) for selected cases.}
\label{marso3spec}
\end{figure}

Considering early Mars as a potential exoplanet, Fig. \ref{marso3spec} shows the resulting emission spectra for five selected cases. They have been calculated with the same high-resolution line-by-line code used for the model verification \citep{schreier2003} and were then binned to a coarse spectral resolution of R=$\frac{\lambda}{\Delta \lambda}$=100. As pointed out by \citet{selsis2002}, the ozone band at 9.6\,$\mu$m coincides with a carbon dioxide hot band and lies in the wings of the strong dimer absorption feature (see, e.g., \citealp{wordsworth2010cont} for the most recent carbon dioxide continuum absorption data). Therefore, except for the strongest ozone column, the shape of the spectral band is not greatly sensitive to atmospheric ozone amount.

\begin{figure}[h]
\centering
  \includegraphics[width=400pt]{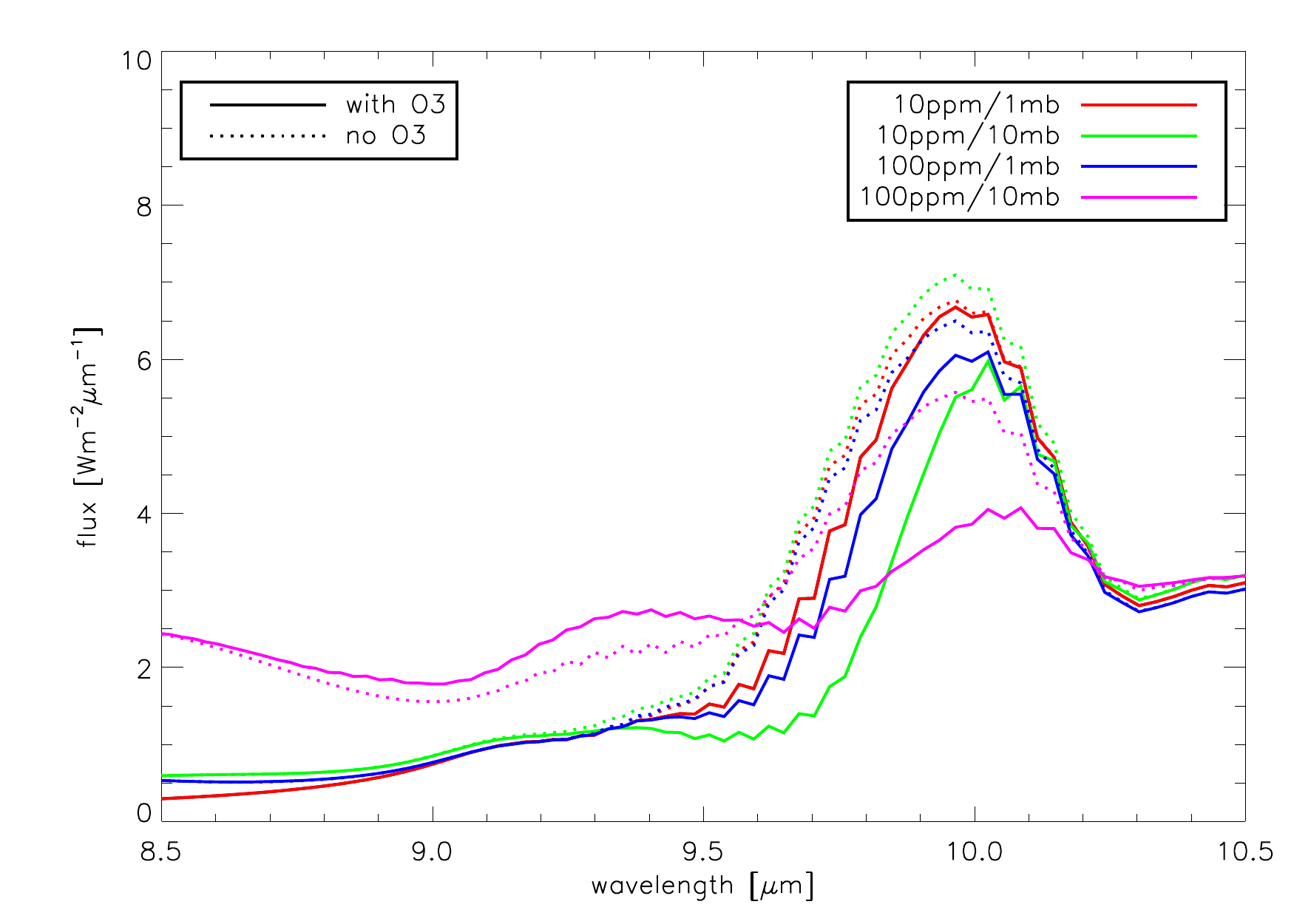}\\
\caption{Zoom in the 9.6\,$\mu$m band of ozone. Effect of including ozone in spectral calculations (R=100).}
\label{marso3zoom}
\end{figure}

To illustrate this, Fig. \ref{marso3zoom} zooms in on the spectral region around the 9.6\,$\mu$m fundamental band of ozone. For comparison, we also show line-by-line calculations with removed ozone (marked  "no O3" in Fig. \ref{marso3zoom}) to investigate the effect of ozone on the spectral appearance and a possible masking by CO$_2$. It is clear that an effect is only seen for the highest ozone columns (higher than the modern-day terrestrial column, Table \ref{columno3}). To detect this, relatively high spectral resolution (here, R=100) and reasonable signal-to-noise ratios (of the order of 4) are needed. These values are not achievable with currently planned instrumentation even for Earth-sized or larger planets, let alone a planet the size of Mars (see, e.g., \citealp{hedelt2013}).

The strong carbon dioxide fundamental band around 15\,$\mu$m is very sensitive to middle-atmosphere temperatures (see Fig. \ref{marso3ir}), and the effect of the changing temperature structure is clearly apparent in this band. In the hypothetical case that all stellar parameters (including UV radiation field) were known, and consistent atmospheric modeling were used, then this band would constitute an excellent indirect probe of atmospheric ozone content. We note, however, that other radiative gases could also be responsible for a temperature inversion (e.g., SO$_2$, CH$_4$, etc.). This then is an important challenge for retrieval algorithms (e.g., \citealp{vparis2013ret}, \citealp{irwin2014}).

In addition to that, the broad 6.3\,$\mu$m band of water, superimposed by part of a carbon dioxide dimer absorption feature, is also sensitive to middle-atmosphere temperatures and thus would be an indirect probe of ozone content. 

Summarizing, in the cases considered for this work, ozone is hardly seen directly in the spectra, except at high ozone contents. Indirectly, however, due to its impact on the thermal structure, the effect of ozone is clearly seen in the spectra.

\begin{figure}[h]
\centering
  \includegraphics[width=300pt]{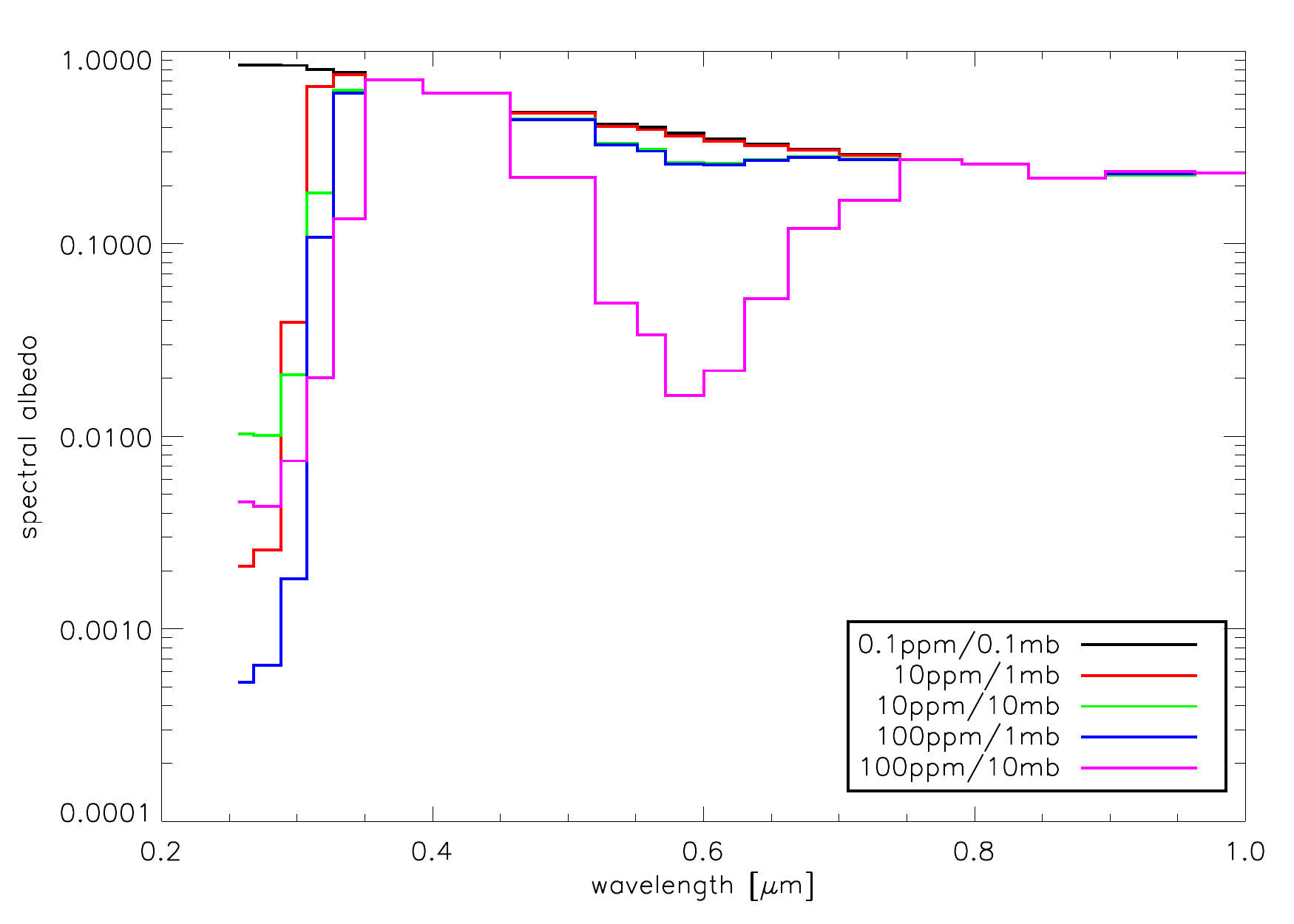}\\
\caption{Effect  of ozone on spectral albedo.}
\label{marso3ref}
\end{figure}

As outlined in \citet{kitzmann2011reflect} or \citet{vparis2013ice}, we use the model output of the stellar radiative transfer code to produce low-resolution albedo spectra. In Fig. \ref{marso3ref}, we show the UV-visible part of the spectrum. For the low-ozone reference case, the spectral albedo clearly follows the Rayleigh scattering slope expected for dense carbon dioxide atmospheres. The presence of ozone leads to a strong reduction of albedo in the Hartley (0.24\,$\mu$m$<\lambda$$<$0.34\,$\mu$m) and Chappuis (0.4\,$\mu$m$<\lambda <$0.7\,$\mu$m) bands. Additionally, the location of the ozone maximum also influences the spectral albedo in the Hartley band, as shown in Fig. \ref{marso3ref}. For cases where the maximum is located at higher pressures, the UV radiation penetrates deeper into the atmosphere and thus is more susceptible to Rayleigh scattering by carbon dioxide.

\section{Conclusions}

\label{summary}

Ozone is a radiatively important atmospheric species due to strong UV absorption bands. It can significantly change the atmospheric temperature structure and thus affect, e.g., cloud formation or atmospheric transport. 

We have investigated the influence of ozone on the atmospheric temperature structure of early Mars. We simulated a 1\,bar carbon dioxide atmosphere with different, fixed ozone concentration profiles. To do this, we have developed a new IR radiative transfer scheme that incorporates molecular absorption of carbon dioxide, water, methane and ozone. Calculated IR cross sections cover a wide range of temperatures, pressures and concentrations. Hence, this scheme can be applied to a wide range of possible planetary scenarios. 

Results  suggest that the impact on surface temperatures at small to moderate ozone concentrations is probably not large, of the order of 1-3\,K, depending on the overall ozone column. The increase in surface temperature is mostly due to the decreased albedo, since ozone strongly absorbs the incoming UV and visible solar radiation. For high ozone concentrations, surface temperatures drop by up to 8\,K due to a change in energy balance. In the upper and middle atmosphere, temperatures increased by up to 60\,K upon introducing ozone. The resulting increase in thermal flux balances the radiative forcing of UV absorption and leads to the observed surface cooling at high ozone concentrations. 

As a consequence of the stratospheric warming, the cloud forming region is reduced. In the case of a thick ozone layer (comparable to or larger than the terrestrial ozone layer), cloud formation is inhibited completely.

For large ozone columns, convection is inhibited. Instead model atmospheres are fully radiative due to a strong reduction of incoming solar flux in the lower atmosphere. This probably decreases the vertical transport.

Upon increasing the ozone content in the model atmospheres, stratospheric water content strongly increased, due to a change in thermal structure and a change in cold trap location. Compared to zero-ozone simulations, water concentrations increased by up to two orders of magnitudes. This is important for assessing possible atmospheric water loss on early Mars, but could be a self-limiting effect due to enhanced catalytic HO$_x$ cycles that destroy ozone.

Future work aims at performing fully coupled climate-chemistry simulations of early Mars as well as carbon dioxide-dominated atmospheres near the outer boundary of the habitable zone.

\section*{Acknowledgements}

This study has received financial support from the French State in the frame of the "Investments for the future" Programme IdEx Bordeaux, reference ANR-10-IDEX-03-02.  This work has been partly supported by the Postdoc Program "Atmospheric dynamics and Photochemistry of Super Earth planets" of the Helmholtz Gemeinschaft (HGF). We thank the two anonymous reviewers for their positive and constructive feedback.

\bibliographystyle{natbib}
\bibliography{literatur_idex}

\end{document}